\documentclass[pra,superscriptaddress,twocolumn,showpacs]{revtex4}%
\usepackage{amsfonts}
\usepackage{amsmath}
\usepackage{amssymb}
\usepackage{graphicx}%
\usepackage{grffile} 
\usepackage{color}

\begin{document}

\title{Electric field control in ultralong-range triatomic polar Rydberg molecules}

\author{M. Mayle}
\affiliation{JILA, University of Colorado and National Institute of Standards and Technology, Boulder, Colorado 80309-0440, USA.}
\author{S. T. Rittenhouse}
\affiliation{ITAMP, Harvard-Smithsonian Center for Astrophysics, Cambridge, MA 02138}
\author{P. Schmelcher}
\affiliation{Zentrum f\"{u}r Optische Quantentechnologien, Universit\"{a}t Hamburg, Luruper
Chaussee 149, D-22761 Hamburg, Germany.}
\author{H. R. Sadeghpour}
\affiliation{ITAMP, Harvard-Smithsonian Center for Astrophysics, Cambridge, MA 02138}

\begin{abstract}
We explore the external electric field control of a species of ultralong-range molecules 
that emerge from the interaction of a ground state polar molecule with a Rydberg atom. The external field mixes
the Rydberg electronic states and therefore strongly alters the electric field seen by the polar diatomic molecule
due to the Rydberg electron. As a consequence, the adiabatic potential energy curves responsible for the molecular binding can
be tuned in such a way that an intersection with neighboring curves occurs. The latter leads to admixture of $s$-wave character in the Rydberg wave function and will substantially facilitate the experimental preparation and realization of this particular class of Rydberg molecule species.
\end{abstract}


\date{\today}
\maketitle

\section{Introduction}
Because of their exaggerated properties (long lifetimes, strong long-range interactions, large sizes, etc.) Rydberg atoms have emerged as leading candidates for studies of strongly correlated systems, many-body quantum phases, quantum information processing, and for non-equilibrium dynamics in cold and ultracold atomic plasmas \cite{Rydoverview}.  A couple of recent developments in the field of cold Rydberg physics are the creation of a frozen Rydberg gas \cite{anderson98}, and the realization of Rydberg blockade and Rydberg qubits \cite{urban09,gaetan09,wilk10,isenhower10}.  One area of particular interest is the prediction and subsequent experimental realization of ultralong-range Rydberg molecules \cite{greene00,bendkowsky09}.  This prototypical giant Rydberg molecule consists of an atom excited to a Rydberg state with large principal quantum number $n$, combined with a ground-state atom that is within the Rydberg electron orbit.  Binding in this system results from frequent, low-energy scattering of the electron off the ground state atom, whose interaction is characterized with a Fermi zero-range potential proportional to the scattering length of Rydberg electron with the ground state atom.  If the scattering length is negative, the zero-range effective interaction is attractive \cite{fermi34} and can bind the ground state atom to the Rydberg atom.  

As with all Rydberg systems, these molecules respond sensitively to external fields \cite{gallagher94,prunele87,hamilton}.  In the original work of Greene {\it et al.}\ \cite{greene00}, a class of Rydberg molecules  (so called ``trilobite'' molecules) were predicted to have massive dipole moments, on the order of kiloDebye (kD), presenting extreme sensitivity to applied electric fields.More recently, an investigation of the Stark spectrum of $s$-wave dominated Rydberg molecules has produced the first direct measurement of a permanent dipole moment in a homonuclear molecule \cite{li11}. The same binding mechanism has been proposed to give rise to ultralong-range giant dipole molecules formed by a neutral alkali ground state atom that is bound in a decentered electronic wave function of a giant dipole atom \cite{kurz2011}.

In this work, we focus on a new type of polyatomic Rydberg molecule formed from the interaction of a Rydberg atom and a polar molecule perturber; first predicted in Ref.~\cite{PhysRevLett.104.243002}.  By considering a nearly degenerate set of higher angular momentum ($l>2$ in rubidium atom) electronic states, it was predicted that a lambda-doublet molecule with a sub-critical Fermi-Teller dipole moment ($d_0<1.639 $ D) can bind to a Rydberg atom. The resulting giant molecules were predicted to be relatively deeply bound, with binding energies $\sim 10$ GHz, and possess very large dipole moments, $d \sim 1$ kD. Furthermore, the molecule can form in two different spatially and energetically separated configurations corresponding to parallel and antiparallel alignment of the perturber's dipole moment. In a followup study \cite{Rittenhouse2011a}, it was found that the electronic structure of the Rydberg molecule could be modified as a function of the perturber polar molecule permanent dipole moment. For large dipole moments ($d_0\gtrsim 1.3$~D for the rubidium Rydberg atoms studied in \cite{Rittenhouse2011a}), a significant amount of $s$-wave character is admixed into the electronic state of the highly localized Rydberg molecule.

The molecules investigated in the present work combine the appealing aspects of ultracold Rydberg and polar molecule systems: the ease of control and manipulation of Rydberg energies and timescales, and the relatively large dipole moments ($\sim 1$ D, as in KRb \cite{ospelkaus10}) of polar molecules in their ground electronic states. Recently, it was shown that state-dependent molecular Rydberg systems could be used to dramatically increase the effective interaction length between polar molecules confined in deep optical lattices. Rydberg atom-mediated interaction with polar molecules provides an additional knob in such systems for single site addressing of molecular qubits  \cite{kuznetsova11}.

In this paper, we examine the behavior of Rydberg molecules formed by a polar perturber with a sub-critical dipole moment under the influence of small external fields ($F_\mathrm{ext}<14$ V/cm).  Due to the existence of large permanent dipole moments for the large triatomic Rydberg molecules, large linear Stark shifts are expected similar to the linear shift for a bare (hydrogen) Rydberg atom \cite{gallagher94}. At zero external field, we confirm that the electronic structure of these molecules has negligible $s$-wave character.  However, due to the strong Stark shift of the molecule, we find that, in the presence of small external fields ($F_\mathrm{ext}\sim 7$ V/cm), a set of broad avoided crossings are introduced in the $s$-wave dominated molecular Born-Oppenheimer (BO) potentials.  This additional $s$-wave coupling induces significant $s$-wave admixture into the molecular electronic state in a manner similar to that seen for larger perturber dipole moments \cite{Rittenhouse2011a}. The sensitivity of these molecules to external fields offers the unique possibility to tune their electronic structure using quite small fields, making them experimentally accessible via standard two photon photo-association \cite{bendkowsky09}.
 
The paper is arranged as follows. In section \ref{sec:theory}, we introduce the adiabatic Hamiltonian which describes the electron-polar molecule interaction considered here, and discuss the methods used to diagonalize it.  In section \ref{sec:convergence}, the convergence behavior of the resulting BO potentials with respect to the Rydberg electron basis set is analyzed.  In section \ref{sec:results}, we present the Stark spectrum and electronic state control of the lowest vibrationally bound states of the Rydberg molecule.  Finally, in section \ref{sec:concl}, we summarize our results and propose future avenues of inquiry.

\section{Adiabatic Hamiltonian}\label{sec:theory}

\begin{figure}
\begin{center}
 \includegraphics[width=7cm]{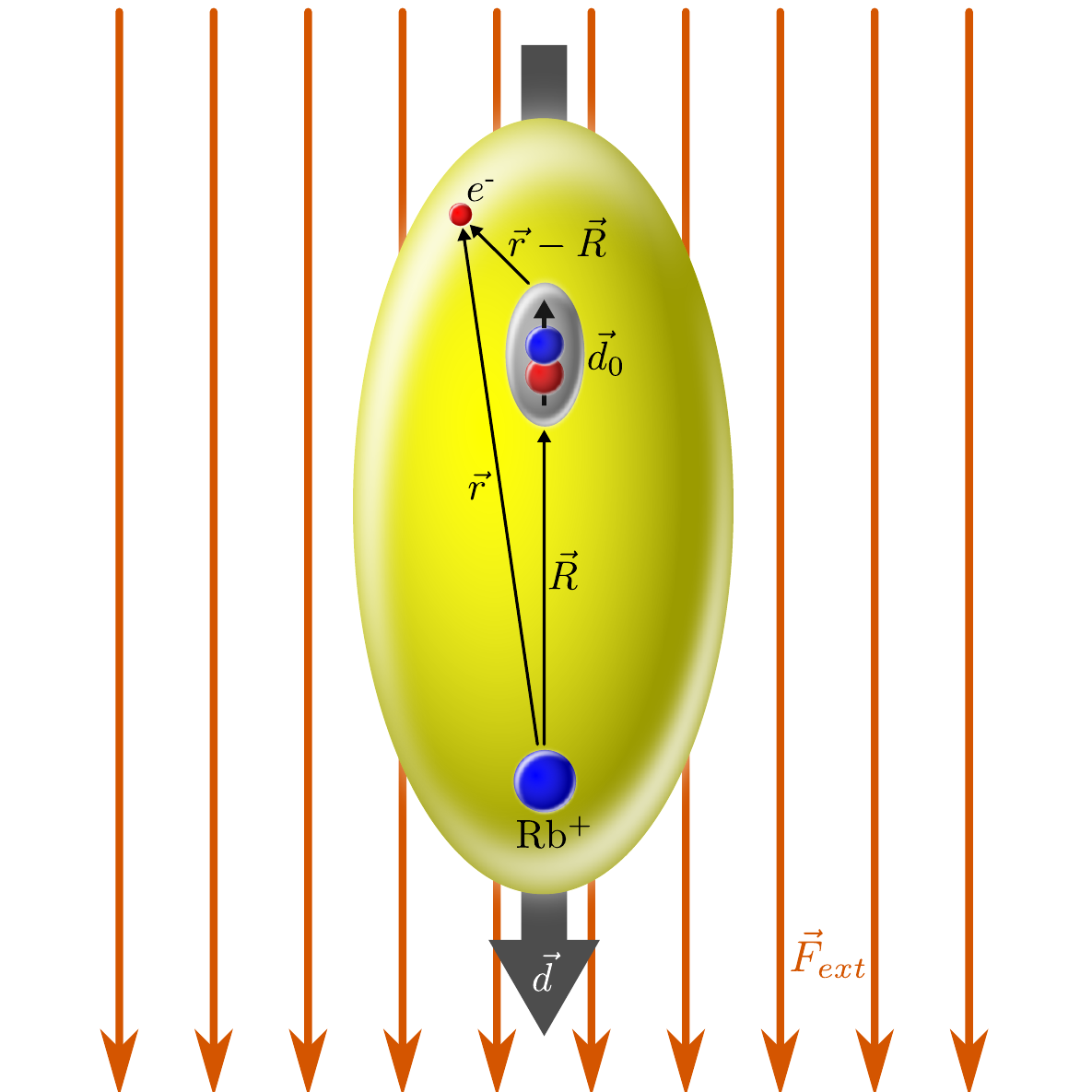}
\end{center}
\caption{A qualitative sketch (not to scale) of the Rydberg atom -- polar perturber system is shown. We adopt the minimal energy configuration where the total dipole $\vec{d}$ is oriented by an external electric field $\vec{F}_\text{ext}$. The dipole $\vec{d}_0$ of the polar perturber, on the other hand, possesses two different parity states, namely, being parallel or antiparallel to the external field, respectively. Here we depict the expected ground state configuration when the dipole resides between the Rydberg electron and the ionic core.
 }%
\label{Fig:diagram}%
\end{figure}

The polar molecule is modeled as a two-level system in which the opposite parity states are mixed in the presence of an electric field, i.e.,
\begin{equation}
H_\text{mol}=
\begin{pmatrix}
0 & -\vec{d}_0\cdot\vec{F}\\
-\vec{d}_0\cdot\vec{F} & \Delta
\end{pmatrix}  , \label{Eq:H_mol}%
\end{equation}
where $\Delta$ is the zero field splitting between the two molecular states (as in a $\Lambda$-doublet molecule), $\vec{d}_0$ is the permanent dipole moment of the molecule in the body fixed frame, and $F$ is an electric field external to the polar molecule ~\cite{Rittenhouse2011a} . Specifically, the electric field stems from the Rydberg electron ($F_e$), the Rydberg ionic core ($F_c$), and a superimposed external field ($F_\text{ext}$),
\begin{align}\label{eq:F}
\vec{F}(\vec{R},\vec{r})  &=
\vec{F}_e+\vec{F}_c+\vec{F}_\text{ext}
\\
&=-e\dfrac{\vec{r}-\vec{R}}{\vert \vec{r}-\vec{R}\vert^{3}}-e \dfrac{\vec{R}}{R^{3}}+\vec{F}_\text{ext} , \label{Eq:Ryd_field}%
\end{align}
where $e$ is the electron charge, $\vec{R}$ is the Rydberg ionic core-polar molecule separation vector and $\vec{r}$ is the position of the Rydberg electron with respect to the core. Eq.~(\ref{Eq:H_mol}) hence describes the coupling between the internal states of the polar molecule and the Rydberg atom, as well as the influence of the external electric field on the polar molecule internal states. The system under consideration is illustrated in Fig.~(\ref{Fig:diagram}).
We remark that, as in Ref.~\cite{Rittenhouse2011a}, the model Hamiltonian (\ref{Eq:H_mol}) is suitable for $\Lambda$ doublet molecules (such as OH and CH) where a permanent dipole arises from the interaction of two opposite parity electronic states. Furthermore, we consider only subcritical dipoles, $d_{0}<d_{c}=1.63$ D \cite{turner:758}.

The Hamiltonian (\ref{Eq:H_mol}) depends on the mutual orientation of the electric fields involved. Here, we assume the minimal energy configuration where $\vec{d}_0$ is aligned by the electric field of the ionic core (defined as the $z$-direction). As shown in Refs.~\cite{PhysRevLett.104.243002,Rittenhouse2011a}, the Rydberg atom -- polar perturber complex gives rise to giant permanent dipole moments $\vec{d}$ in the kD range. A small external field will immediately orient the dipole along the field direction. Here we choose the external field to be antiparallel to the $z$-axis, $\vec{F}_\text{ext}=-F_\text{ext}\vec{e}_z$, such that the configuration depicted in Fig.~\ref{Fig:diagram} results in which the polar perturber is localized on the positive $z$-axis, $\vec{R}=R\hat{z}$. Eq.~(\ref{eq:F}) then becomes
\begin{align}
F(\vec{R},\vec{r})  =\dfrac{e\cos\theta_{\vec{r}-\vec{R}}}{\vert \vec{r}-\vec{R}\vert^{2}}+\dfrac{e}{R^{2}}-F_\text{ext},
\label{Eq:zaxis_field}%
\end{align}
where $\vec{F}(\vec{R},\vec{r})=F(\vec{R},\vec{r})\vec{e}_z$ and $\theta_{\vec{r}-\vec{R}}=(\vec{r}-\vec{R})\cdot\vec{R}/R\vert\vec{r}-\vec{R}\vert$. This means that the projection $m$ of the Rydberg electron angular momentum along $\vec{R}$ is conserved. 

To find the BO potentials, we solve the adiabatic Schr\"{o}dinger equation at fixed polar molecule location $\vec{R}=R\hat{z}$,
\begin{align}
H_\text{ad}\psi\left(  R;\vec{r},\sigma\right)   &  =U(R)  \psi(
R;\vec{r},\sigma)  ,\label{Eq:Adiab_SE}
\end{align}
with
\begin{equation}\label{eq:Had}
H_\text{ad}   = H_A +H_\text{mol}.
\end{equation}
As in Ref.~\cite{Rittenhouse2011a}, the first term, $H_A=-\frac{\hbar^{2}}{2m_{e}}\nabla_{r}^{2}+V_l(r)+H_S$, describes the unperturbed Rydberg atom but now with addition of the Stark effect 
\begin{equation}
H_S=F_\text{ext}r\cos\theta_{\vec{r}} 
\end{equation}
of the Rydberg atom. The core penetration, scattering, and polarization effects of the Rydberg electron are accounted for by the $l$-dependent model potential $V_l(r)$ \cite{PhysRevA.49.982}, giving rise to the quantum defects of the low angular momentum Rydberg states. $m_{e}$ is the electron mass, $\psi$ is the electron wave function, and $\sigma$ is a collective coordinate for the internal states of the polar molecule. The eigenvalues $U(R)$ of Eq.~(\ref{Eq:Adiab_SE}) serve as the BO potentials for the Rydberg molecule. 

To solve Eq.~(\ref{Eq:Adiab_SE}), the total wave function is expanded in the basis $\{\psi_{nlm}\left(\vec{r}\right) \left\vert \pm\right\rangle\}$ where $\psi_{nlm}\left(  \vec{r}\right)$ is an unperturbed Rydberg orbital and $\left\vert \pm \right\rangle $ are the polar molecule parity states.  For low angular momentum states ($l\le2$), where the core effects matter, the Rydberg orbitals are obtained by solving the field-free atomic Hamiltonian $H_A$ by means of a discrete variable technique \cite{mayle:053410}. For high angular momentum states ($l>2$) the Rydberg orbitals are taken as the hydrogenic wave functions. The Hamiltonian (\ref{eq:Had}) is diagonalized in an extended basis set comprising several Rydberg $n$ manifolds (see Sec.~\ref{sec:convergence} for a discussion on the actual number of Rydberg manifolds needed to achieve convergence). The matrix elements for $F_\text{ext}=0$ are presented in Ref.~\cite{Rittenhouse2011a} and require the numerical integration of highly oscillating functions. The matrix elements of the atomic Stark Hamiltonian $H_S$ need either to be determined numerically whenever a low angular momentum state is involved or can be calculated analytically for the hydrogenic eigenstates,
\begin{align}
 \langle \psi_{n^{\prime}l^{\prime}m^{\prime}},\sigma^{\prime}\vert H_S\vert \psi_{nlm},\sigma\rangle
={}&\delta_{\sigma\sigma'}\delta_{mm'}\mathcal{A}_{l'm,lm}\mathcal{R}_{n'l',nl},
\end{align}
with
\begin{widetext}
\begin{align}
\mathcal{A}_{l'm,lm}={}&\sqrt{\frac{(l+m)(l-m)}{4{l}^2-1}}\,\delta_{l'l-1}+\sqrt{\frac{(l+m+1)(l-m+1)}{(2l+1)(2l+3)}}\,\delta_{l'l+1},\\
 \mathcal{R}_{n'l',nl}={}&2^{l+l'+2}n^{-l-2}{n'}^{-l'-2}\sqrt{\frac{(n-l-1)!(n'-l'-1)!}{(n+l)!(n'+l')!}}\\
&\times
\sum_{m=0}^{n-l-1}\sum_{m'=0}^{n'-l'-1}(-1)^{m+m'}
\begin{pmatrix}
 n+l\\n-l-1-m
\end{pmatrix}
\begin{pmatrix}
 n'+l'\\n'-l'-1-m'
\end{pmatrix}\frac{2^{m+m'}}{m!m'!}\frac{1}{n^m{n'}^{m'}}\frac{(3+l+l'+m+m')!}{(\frac{1}{n}+\frac{1}{n'})^{4+l+l'+m+m'}}\,.
\end{align}
\end{widetext}

Overall, the adiabatic Hamiltonian matrix assumes the following form,
\begin{equation}
\bar{H}_\text{ad}=\left(
\begin{array}
[c]{cc}%
\bar{H}_A & d_{0}\bar{F}\\
d_{0}\bar{F} & \Delta\bar{I}+\bar{H}_A%
\end{array}
\right),  \label{Eq:Had_mat}%
\end{equation}
where $\bar{I}$ is the unity matrix, $\bar{H}_A=\bar{E}_{Ryd}+\bar{H}_{S}$ is the matrix representation of $H_A$ with $\bar{E}_{Ryd}$ being the diagonal matrix containing the field-free Rydberg energies, and $\bar{F}$ is the electric field matrix whose elements are given in \cite{Rittenhouse2011a} with an additional diagonal offset of $F_\text{ext}$ due to the external electric field. The atomic and molecular degrees of freedom are coupled by the electric field (\ref{Eq:zaxis_field}) that mixes both the Rydberg orbitals as well as the parity states of the polar molecule, cf.\ Eq.~(\ref{Eq:H_mol}). The BO potentials are found by diagonalizing the matrix (\ref{Eq:Had_mat}) at each $R$. The quantum defects used in this work for the determination of $\bar{E}_{Ryd}$ are those for the rubidium atom \cite{gallagher94}: $\mu_{s}=3.13$, $\mu_{p}=2.65$, $\mu_{d}=1.35$, $\mu_{f}=0.016$, and $\mu_{l>3} \approx0$. While our focus is on Rb, the current technique can be extended to any highly excited Rydberg atom by using the appropriate quantum defects.

\section{Basis Set}\label{sec:convergence}

\begin{figure}
 \includegraphics[width=1\columnwidth]{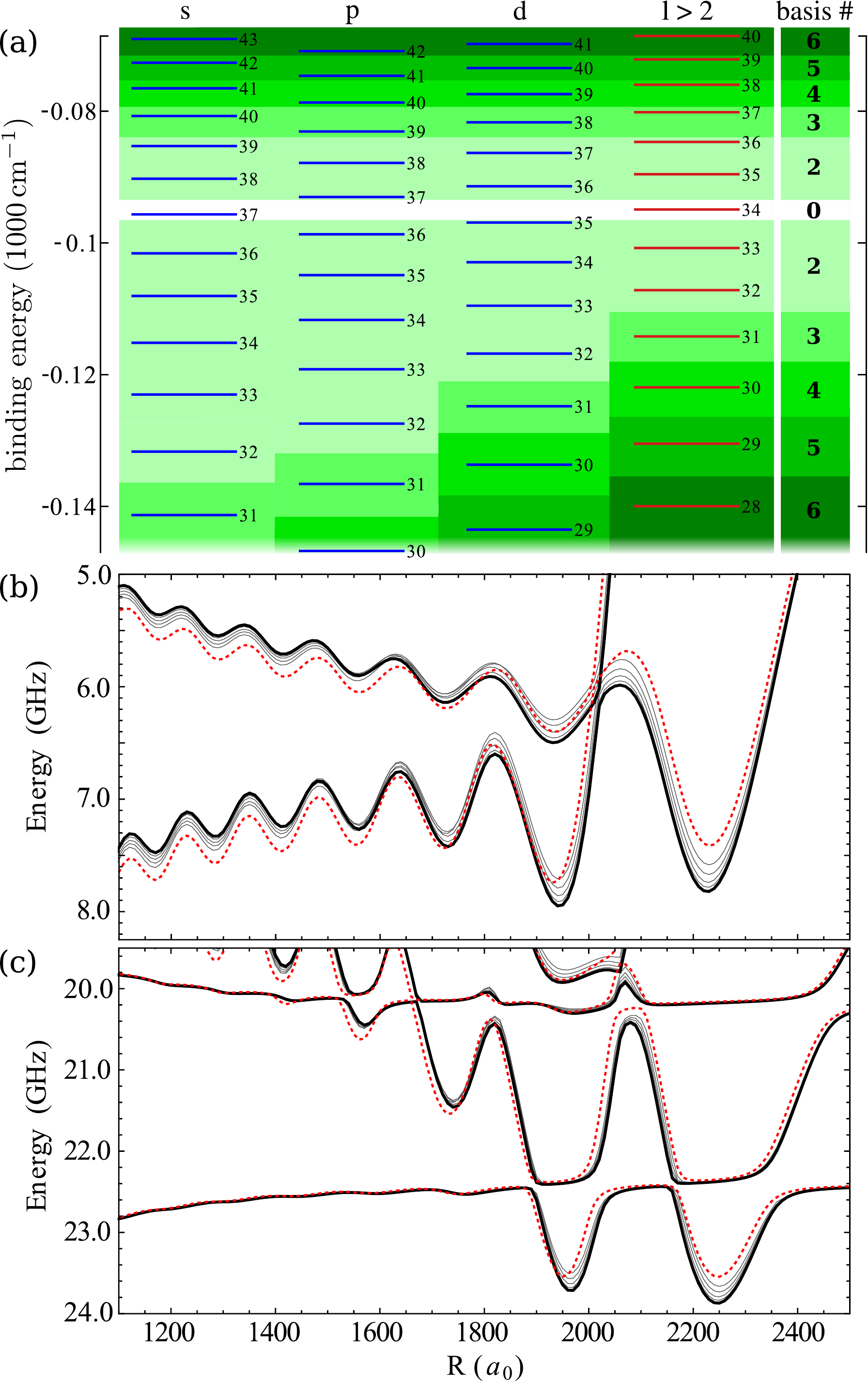}
\caption{(color online)
 (a) Illustration of the basis sets used for calculating the potential energy surfaces stemming from the $n=34$ Rydberg manifold. The zero order basis includes the $\{n=34,l>2\}$ manifold as well as the $37s$ quantum defect split state, energetically close to the $n=34$ manifold; the Stark effect of the $37$s state has been included. The larger basis sets include the $n-\Delta n,\dots,n+\Delta n$ manifolds as well as the $\{(n+\Delta n+1,s,p,d),(n+\Delta n+2,s,p),(n+\Delta n,s)\}$ states and are labeled by $\Delta n$. For all the basis sets, only the $m_l=0$ is considered. (b,c) Convergence of the potential energy surfaces with respect to the basis set for zero external electric field and for $F_\text{ext}=8.0\,$V/cm, respectively. In both panels, the solid black line corresponds to the largest basis set considered, $\Delta n=6$; the solid gray lines are for decreasing $\Delta n=5,4,3,2$. The red dashed line illustrates the zero-order results, for which already a good agreement with the $\Delta n=6$ result can be found.
\label{fig:convergence}
}
\end{figure}

In calculating the BO potential surfaces -- by diagonalizing Hamiltonian (\ref{Eq:Had_mat}) -- care must be exercised with respect to the basis size. The larger the dipole $d_0$ of the polar perturber, the more Rydberg orbitals are needed to describe the localization of the Rydberg electron at the dipole. In this work, we consider a model perturber whose dipole moment corresponds to that of a rigid rotor molecule (KRb) in its ground rovibrational state, $d_0=0.566$ D \cite{Ni2008a}. Accordingly, the splitting $\Delta$ is chosen as the rotational splitting of the KRb molecule between the ground and first excited rotational state, $\Delta=2.228$ GHz \cite{PhysRevLett.104.030402}. As mentioned before, we remark that for non-$\Lambda$-doublet molecules such as KRb, the present framework is not suitable. Hence, while we do not expect quantitative predictions for the KRb molecule, significant insight can be gained for $\Lambda$ doublet molecules with subcritical dipoles of the same order as that for KRb.

In Fig.~\ref{fig:convergence}, we provide a convergence study of the BO potentials stemming from the $n=34$ Rydberg manifold as a function of the basis size. Fig. \ref{fig:convergence}(a) illustrates the basis: The zero order basis (highlighted in white) includes the $\{n=34,l>2\}$ manifold as well as the $37s$ quantum defect split state which is energetically close to the $n=34$ manifold. The larger basis sets include an increasing number $\Delta n$ of Rydberg manifolds, distributed symmetrically around the intended Rydberg state $n$ (highlighted in increasingly darker shades of green/gray); in addition the $\{(n+\Delta n+1,s,p,d),(n+\Delta n+2,s,p),(n+\Delta n,s)\}$ states are included. For all bases, only $m_l=0$ is considered. In Fig. \ref{fig:convergence}(b), the BO potentials for the different basis sets for $F_\text{ext}=0$ are displayed. A clear convergence trend from the zero order basis (red dashed line, 32 basis functions) to the largest basis set ($\Delta n=6$, black solid line, 448 basis functions) is found. The zero order basis already provides quantitatively adequate results. Hence, we will use the zero order calculations as the basis for the ensuing discussions.

It is not \emph{a priori}  clear if the zero order basis is able to produce satisfactory results for the field-dressed BO potentials. In Fig. \ref{fig:convergence}({c}), the convergence study is presented for an external field $F_\text{ext}=8.0\,$V/cm which is chosen such that the BO surfaces cross through the $37s$ threshold. One finds basically the same convergence behavior as in the field-free case, validating the use of the zero order basis also for finite external fields. We remark that even in the zero order basis we accounted for the small quadratic Stark effect of the $37s$ state. The two lowest asymptotes are split in by the parity splitting, $\Delta=2.228$ GHz.

\section{Results}\label{sec:results}

\begin{figure}
 \includegraphics[width=\columnwidth]{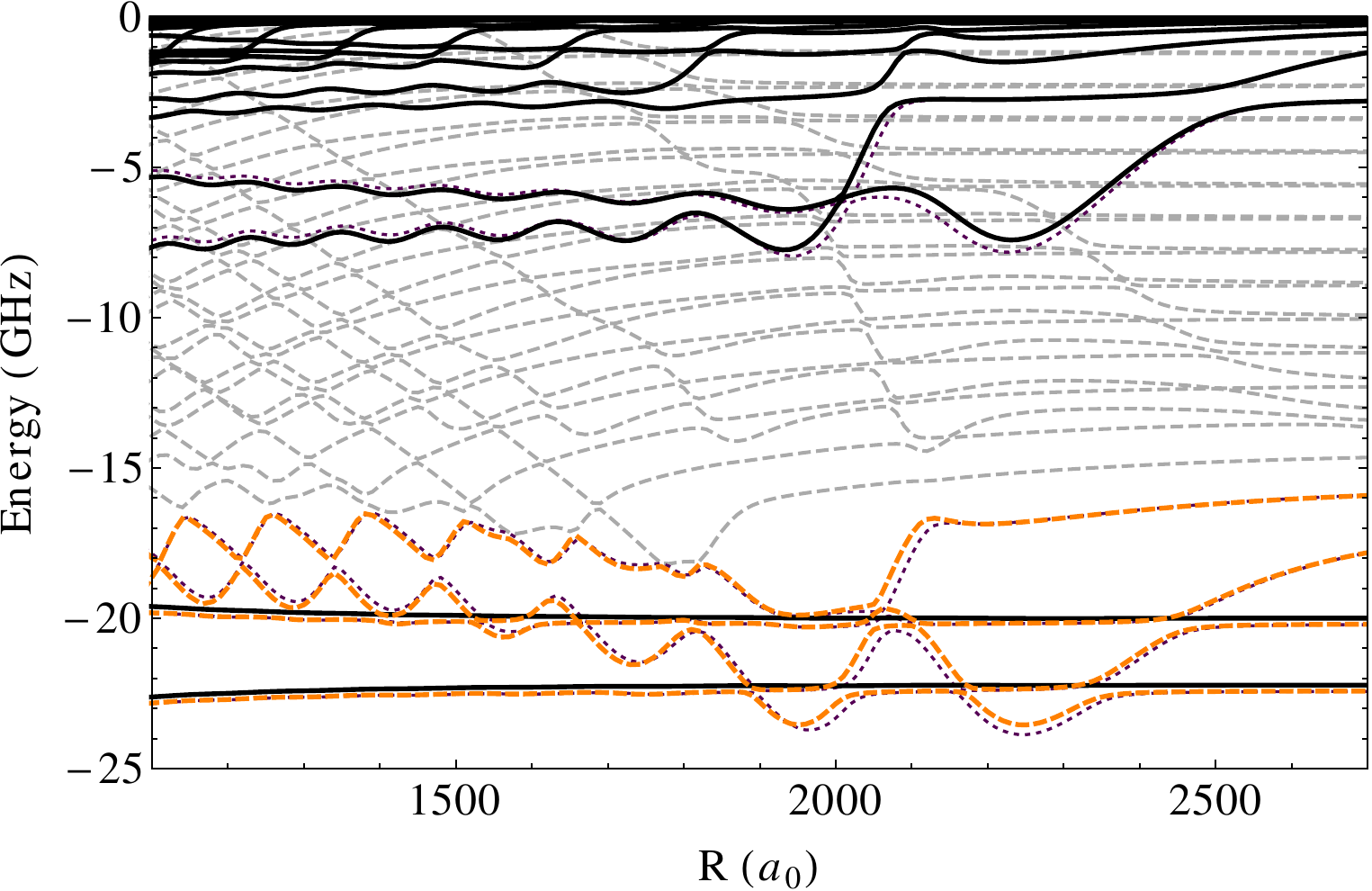}
\caption{(color online)
Overview of the potential energy surfaces resulting from the interaction of a $n=34$ Rydberg atom with a polar perturber of strength $d_0=0.566$ Debye and a parity state splitting of $\Delta=2.228$ GHz. Solid lines: zero external field; dashed lines: $F_\text{ext}=8.0\,$V/cm. The energy surfaces are calculated using the zero order basis; the dotted lines indicate the results from a calculation using the $\Delta n=6$ basis set for the lowest three energy surfaces. The two lowest asymptotes are split in by the parity splitting, $\Delta$.
\label{fig:overview}
}
\end{figure}

The most salient features of the BO potentials in the field-free case were already discussed in Refs.~\cite{PhysRevLett.104.243002,Rittenhouse2011a}: The modulations that form a series of wells reflect the oscillatory nature of the Rydberg electron wave function and the outer most wells in the lowest two
potentials are deep enough to support many vibrational levels. The size of the resulting polyatomic
Rydberg molecules scales as $R_{ryd}\propto n^{2}$ and the well depths scale as
$V_{D}\propto d_{0}/n^{3}$.

\begin{figure}
 \includegraphics[width=0.96\columnwidth]{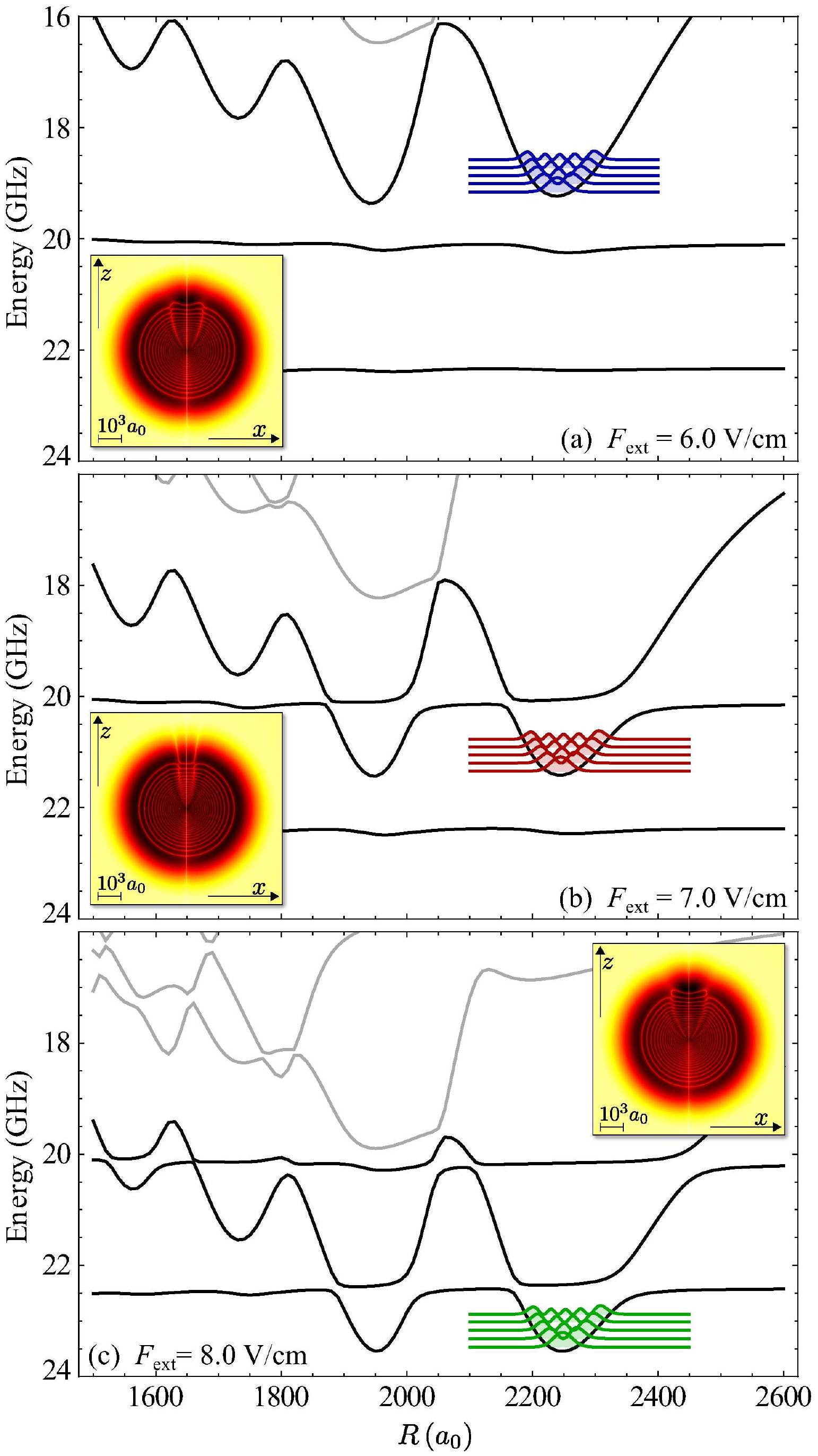}
\caption{(color online)
The BO potentials for the Rydberg-polar molecule system are shown, calculated for Rb$(n=34)$ using the zero order basis and the molecular parameters $d_0=0.566$ Debye, $\Delta=2.228$ GHz. The external electric field strengths considered are (a) $F_\text{ext}=6.0$ V/cm, (b) $F_\text{ext}=7.0$ V/cm, and (c) $F_\text{ext}=8.0$ V/cm. In the outermost potential wells the modulus squared of the first few vibrational wave functions are indicated, calculated assuming the mass of the KRb molecule for the polar perturber. The insets show density plots of the electronic wave function that gives rise to the BO potential where the vibrational states are localized; the wave functions are depicted in cylindrical coordinates and for a fixed polar perturber position of $R=2250\,a_0$ which corresponds to the minimum of the outermost well. 
\label{fig:wave}
}
\end{figure}

The addition of an external electric field has potentially two effects: it can act directly on the molecular dipole $\vec{d}_0$ and it can influence the Rydberg atom. The former is negligible since the electric field due to the ionic core dominates in the range of energies we are interested in. The Rydberg atom, on the other hand, is highly susceptible to external fields. Moreover, the Stark effect is state dependent: While the degenerate $l>2$ manifold shows a strong linear Stark effect, the low angular momentum states, which are split by the quantum defect, possess weak quadratic Stark effect. In particular, the $(n+3)s$ state, which is the energetically closest state to the degenerate $n$ manifold, shows an almost constant behavior for small fields. As a result, the $n(l>2)$ levels cross through the $(n+3)s$ state already for small electric fields ($F\approx7-8\,$V/cm for $n=34$). This provides an excellent handle for the field-controlled manipulation of the BO molecular states.

An overview of the relevant BO surfaces for two different external fields, namely, $F_\text{ext}=0$ and $F_\text{ext}=8.0\,$V/cm, is provided in Fig.~\ref{fig:overview} for the $n=34$ polyatomic Rydberg molecules. At zero field (solid lines), the $n=34(l>2)$ BO potentials are well separated from the quantum defect split $37s$ state. As indicated above, the external field results in a linear shift of the degenerate states while the $37s$ state is barely affected. Accordingly, already for a small external field of $F_\text{ext}=8.0\,$V/cm (dashed lines) the outer two wells of the $n=34$ surfaces are no longer separated from the $37s$ state but cut through it. In the absence of an external field, a similar effect can be achieved if the molecular dipole $d_0$ is closer to its critical value \cite{Rittenhouse2011a}. 

\begin{figure}
 \includegraphics[width=\columnwidth]{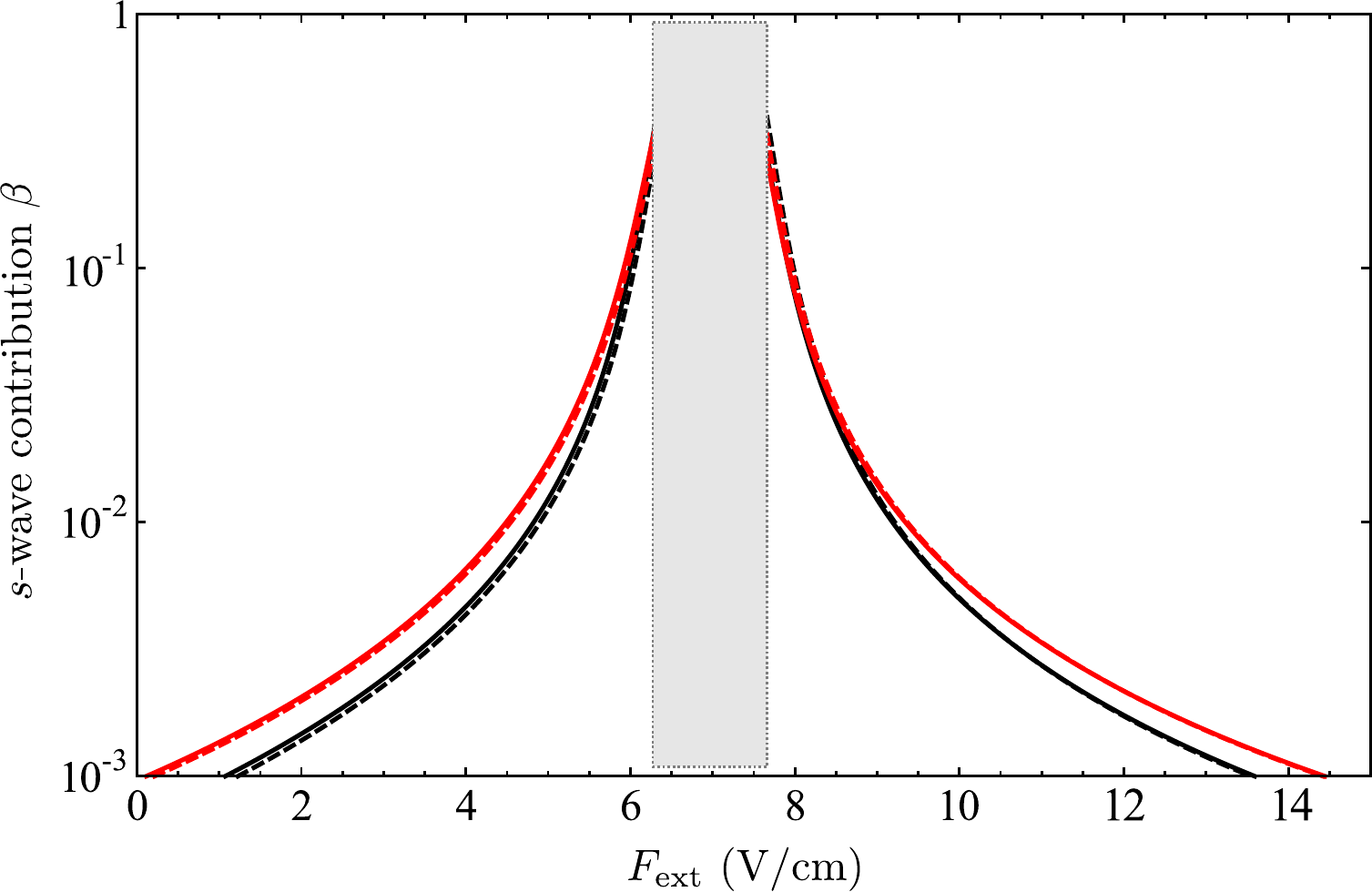}
\caption{(color online)
$s$-wave character $\beta$, cf.\ Eq.~(\ref{eq:beta}), of the first (solid lines) and second (dashed lines) vibrational states of the right (black) and left (red/gray) outer well; the first and second vibrational states show a very similar behavior and hence partially overlap. The zero order basis set has been used for the determination of $\beta$. The gray shaded area corresponds to the regime of field strengths where the wells cross through the $37s$ states and hence a determination of the vibrational states in a single-channel picture.
\label{fig:swave}
}
\end{figure}

A closer look at the transition of the potential wells through the $37s$ threshold is provided in Fig.~\ref{fig:wave} where closeups of the BO potentials for the three field strengths (a) $F_\text{ext}=6.0$\,V/cm, (b) $F_\text{ext}=7.0$\,V/cm, and (c) $F_\text{ext}=8.0$\,V/cm are presented. The first few vibrational states supported by the outer well are shown at their respective energies; note that all vibrational states are computed based on purely adiabatic potentials, neglecting nonadiabatic couplings. The potentials support multiple vibrational states \cite{PhysRevLett.104.243002,Rittenhouse2011a}. This holds even in the case where the $n=34$ potential wells are strongly mixed with the $37s$ states, cf.\ Fig.~\ref{fig:wave}(b). In this case, the potentials acquire a substantial amount of $s$-wave character thereby, in principle, allowing for the production of the polyatomic Rydberg molecules by two-photon excitation from a mixture of ultracold ground state atoms and molecules; see also Fig.~\ref{fig:swave} for a further discussion of the $s$-wave admixture.

The influence of the $s$-wave states can be also observed in the electronic wave functions at the minimum of the potential wells, as shown in the insets of Fig.~\ref{fig:wave}. Just before and after the potentials cut through the $37s$ threshold, cf.\ Figs.~\ref{fig:wave}(a,c), one finds electronic wave functions typical for the polyatomic Rydberg molecules based on the high angular momentum Rydberg states \cite{PhysRevLett.104.243002}. For the case of Fig.~\ref{fig:wave}(b), however, the strong $s$-wave influence diminishes the localization of the Rydberg electron around the polar perturber and the wave function starts to approach the spherical symmetry of the $s$-wave state.

\begin{figure}
\begin{flushright}
\includegraphics[width=0.9825\columnwidth]{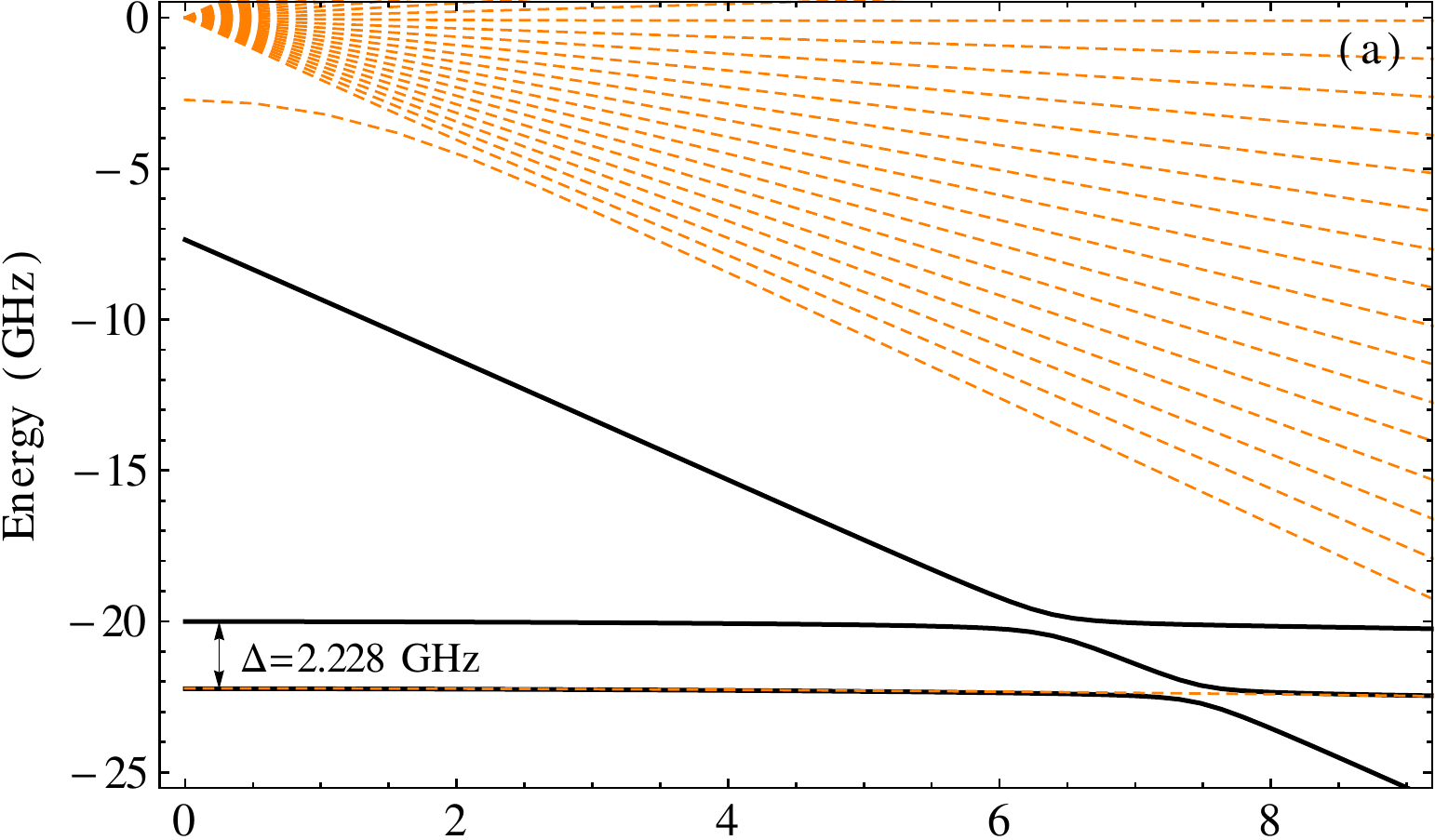}\vspace{0.1cm}\\
\includegraphics[width=0.99\columnwidth]{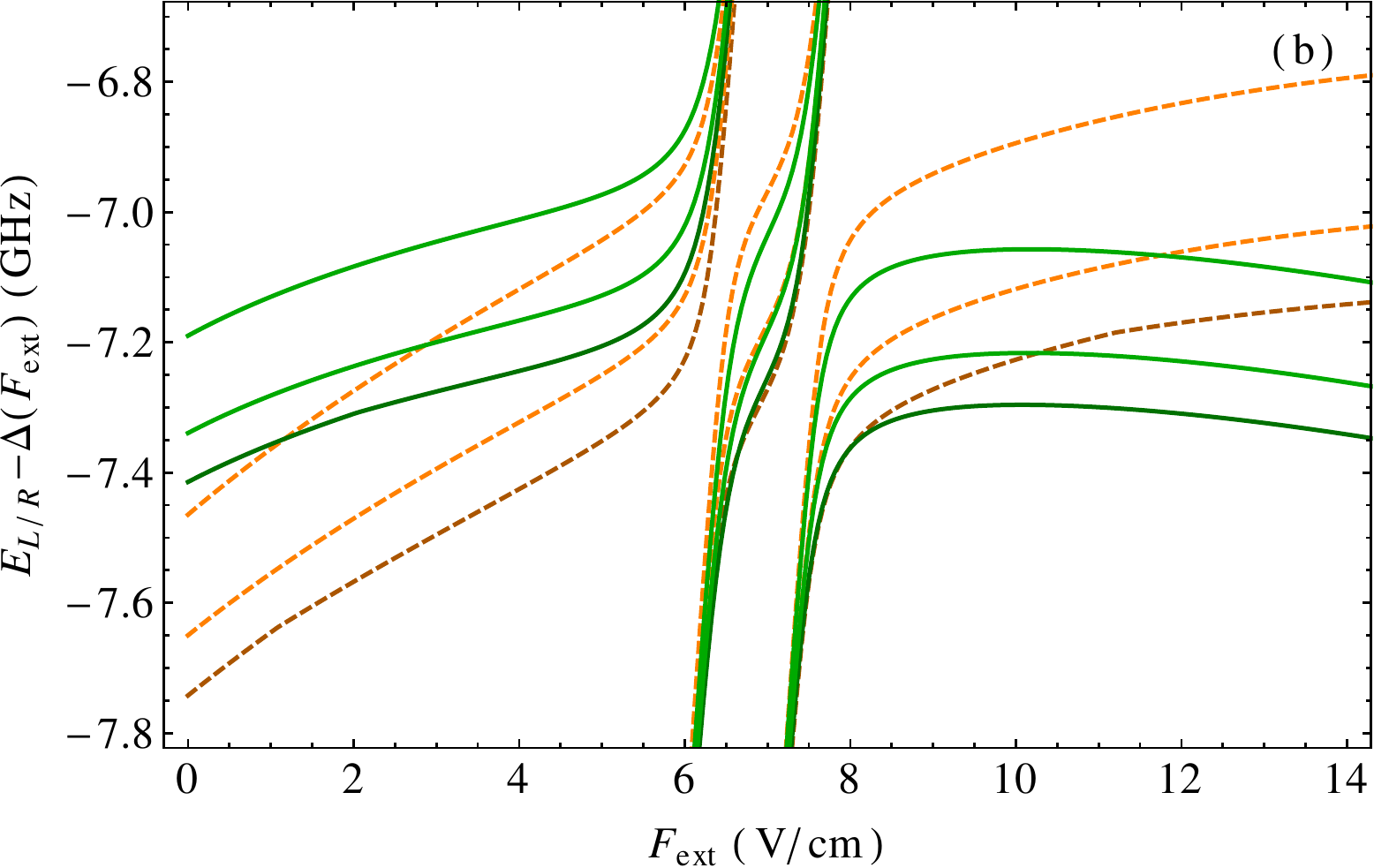}
\end{flushright}
\caption{(color online)
(a) Stark effect $E_R$ of the outermost potential well (solid line) which follows the Stark effect of the unperturbed $n=34,l>2$ Rb Rydberg manifold (dashed lines); for the latter, the small $f$-wave quantum defect is included. Panel (b) shows the Stark effect of the outermost two potential wells (solid: $E_R$/right well, dashed: $E_L$/left well), as well as of the first two vibrational states supported by these wells, offset by the linear Stark effect Eq.~(\ref{eq:offset}) of the $n=34,l>2$ manifold. The parameters ($n=34$, $d_0=0.566$ Debye, and $\Delta=2.228$ GHz) are taken to mimic a KRb molecule where the parity splitting corresponds to the splitting of the first and second rotational states; accordingly, the KRb reduced mass has been used for the determination of the vibrational states. 
\label{fig:stark}
}
\end{figure}

A more detailed investigation of the Stark effect of the potentials is found in Fig.~\ref{fig:stark}. The subfigure \ref{fig:stark}(a) shows the Stark effect of the outermost potential well (solid lines), i.e., the field dependent energy $E_R$ of the associated minimum. The Stark effect of the Rydberg states is indicated as orange/gray dashed lines. Because the right well stems from the lowest Stark state of the $n=34,l>2$ manifold, $E_R$ shows a linear Stark shift of approximately
\begin{equation}\label{eq:offset}
\Delta(F_\mathrm{ext})=-\frac{3}{2} n (n - 3) F_\mathrm{ext}.
\end{equation}
Deviations from Eq.~(\ref{eq:offset}) are due to the fact that we include the $f$-wave quantum defect. As can also be seen in Fig.~\ref{fig:stark}(a), the Stark shift of the $37$s Rydberg state is negligible on the scale of energies considered here, allowing for a crossing of the potential wells through the $s$-wave thresholds already for relatively small external fields of about $F_\mathrm{ext}\approx7$ kV/cm.

Fig. \ref{fig:stark}(b) shows the Stark effect of the two outermost potential wells, $E_R$ and $E_L$, offset by 
$\Delta(F_\mathrm{ext})$. Since the latter corresponds approximately to the linear Stark effect of $E_{R/L}$, one finds an almost constant behavior as a function of $F_\mathrm{ext}$. In addition, one clearly observes the deviation from the linear Stark effect when the potential wells cross through the $s$-wave threshold. The non-linear behavior near $F_\text{ext}\approx7\,$V/cm occurs after subtracting the linear offset (\ref{eq:offset}), revealing the strong admixture of the $s$-wave state.
In addition to the Stark effect of the BO surface, Fig.~\ref{fig:stark}(b) features the actual energies of the first two vibrational states of either of the two outermost potential wells as a function of the external field $F_\text{ext}$. As expected, they follow the Stark effect of the corresponding BO surfaces. However, as the wells approach the $s$-wave threshold, strong avoided crossings flatten the potential wells, leading to smaller vibrational splittings.

More details on the $s$-wave character of the polyatomic Rydberg states can be found in Fig.~\ref{fig:swave}. We define the $s$-wave contribution of a certain polyatomic Rydberg state as
\begin{equation}\label{eq:beta}
 \beta=\langle \chi(R)|b(R)^2|\chi(R)\rangle,
\end{equation}
where $\chi(R)$ is its vibrational wave function and $b(R)$ defines the $s$-wave admixture of the electronic wave function for a fixed position $R$ of the polar perturber via 
\begin{equation}\label{Eq:Mol_wf_mixture}
\psi( R;\vec{r})  =a(R) \psi_{d}(\vec{r}) + b(R) \psi_{s}(\vec{r});
\end{equation}
$\psi( R;\vec{r})$ denotes the electronic part of the adiabatic channel function with $\psi_{d}(\vec{r})$ including the higher electronic angular momentum Rydberg states with $l>2$ and $\psi_{s}(\vec{r})$ being the $s$-wave Rydberg electron wave function. As expected, $\beta$ peaks for external fields where the potentials cross through the $37s$ threshold, which occurs rather rapidly. Hence, the admixing of large amounts of $s$-wave character $(\beta \gtrsim 20$\%) becomes possible for initially near-degenerate high-$l$ polyatomic Rydberg molecules via the application of a small external electric field. We excluded in Fig.~\ref{fig:swave} the region where the potentials wells $E_{R/L}$ are in between the two $s$-wave thresholds (split by the $\Lambda$-doublet splitting), as the determination of the vibrational states in this region is no longer characterized in the single-channel picture, due to multiple avoided crossings between the adiabatic channels.

\section{Conclusions and outlook}\label{sec:concl}

In this paper, we have described the behavior of giant, polyatomic, Rydberg molecules formed by a polar perturber in the presence of weak external electric fields. For smaller dipole moments ($d \sim 0.5$ D) we showed that the BO potentials are converged using a surprisingly small set of Rydberg orbitals.  Using only a $n(l>2)$ nearly degenerate set of hydrogenic Rydberg orbitals, augmented with the $(n+3)s$ state, the BO potentials are converged to within approximately 10\% at zero field.  Similar convergence behavior was found at finite fields as long as the quadratic Stark shift on the $(n+3)s$ state was taken into account. 

Many of the salient features of these molecules persist in the presence of an external field, most importantly their large size ($R_0\approx 2000\text{ a.u.} \sim n^2$), and an isolated double well structure which corresponds to different orientations of the polar perturber dipole moment. As expected, these molecules exhibit a strong linear Stark shift at small external fields due to their extremely large dipole moments, $d \approx 4$ kD for $n=34$.

Due to their extreme sensitivity to external fields, small fields can be used to significantly influence the electronic structure of these molecules. For a small range of external fields ($6.0 \text{ V/cm}<F_\mathrm{ext}<8.0 \text{ V/cm}$ for the $n=34$ states studied here), the outer-most potential wells go through a set of avoided crossings with two $s$-wave dominated potentials.  These avoided crossings significant modify the structure of the Rydberg electron wave functions, admixing large amounts of $s$-wave character $(\beta \gtrsim 20$\%).  For larger fields, the wells pass through the $s$-wave threshold, and the electronic structure corresponding to their minima reverts to highly-localized behavior.   The avoided crossings should change the dynamics and lifetimes of the molecule.  These topics are the subjects of ongoing studies.

We have focused on molecules formed from a rubidium Rydberg atom perturbed by a $\Lambda$-doublet molecule whose dipole moment and zero-field splitting were chosen to agree with the dipole moment, and rotational splitting of KRb.  We expect that the behavior demonstrated here will persist, however, for any system consisting of an alkali metal Rydberg atom and a $\Lambda$-doublet molecule with a sufficiently small (subcritical) dipole moment ($d_0\lesssim 0.6$ D), though the field value at which $s$-wave avoided crossings are induced will depend on the principal quantum number, the atomic species, and the perturber dipole moment.  While the predictions here are not likely to be applicable to more conventional rigid-rotor type polar molecules, we can expect that Rydberg molecules formed using such molecules will exhibit similar topological features (e.g., a double well structure and field dependent $s$-wave admixtures).  A more realistic model of the polar perturber for this type of system is needed and is a subject of ongoing inquiry.

\section{Acknowledgments}
MM acknowledges financial support by a fellowship within the postdoc-programme of the German Academic Exchange Service (DAAD).
STR and HRS acknowledge financial support from the NSF through ITAMP 
at Harvard University and the Smithsonian Astrophysical Observatory.

\end{document}